\documentclass[sigconf, nonacm]{acmart}

\usepackage[many]{tcolorbox}    	

\definecolor{main}{HTML}{fcdc50}    
\definecolor{sub}{HTML}{fcf2ba}     

\usepackage{dirtytalk}
\usepackage{csquotes}

\usepackage[inkscapelatex=false]{svg}

\graphicspath{ {./img/} }
       
\newenvironment{myparticipantquote}[2]
{\say{\textit{#1}} (#2)}

\acmDOI{XXXXXXX.XXXXXXX}
\acmConference[EASE 2026]{The 30th International Conference on Evaluation and Assessment in Software Engineering}{9–12 June, 2026}{Glasgow, Scotland, United Kingdom}

\acmISBN{978-1-4503-XXXX-X/2018/06}

\begin{document}

\title[Exploring Creativity in Human-Human-LLM Collaborative Software Design]{Exploring Creativity in Human–Human-LLM Collaborative Software Design}

\author{Victoria Jackson}
\email{v.jackson@soton.ac.uk}
\orcid{0000-0002-6326-931X}
\affiliation{%
  \institution{University of Southampton}
 \city{Southampton}
  \country{UK}
}

\author{Grischa Liebel}
\email{grischal@ru.is}
\orcid{0000-0002-3884-815X}
\affiliation{%
  \institution{Reykjavik University}
 \city{Reykjavik}
  \country{Iceland}}

\author{Rafael Prikladnicki}
\email{rafaelp@pucrs.br}
\orcid{0000-0003-3351-4916}
\affiliation{%
  \institution{Pontifícia Universidade do Rio Grande do Sul}
 \city{Porto Alegre, RS}
  \country{Brazil}}

\author{Andr\'e van der Hoek}
\email{andre@ics.uci.edu}
\orcid{0000-0001-7917-932X}
\affiliation{%
  \institution{University of California, Irvine}
 \city{Irvine}
  \country{USA}
}

\renewcommand{\shortauthors}{Jackson, Liebel, Prikladnicki, van der Hoek}

\begin{abstract}
  While the use of Large Language Models (LLMs) in programming has been extensively studied, there is limited understanding of how LLMs support collaborative work where creativity plays a central role. 
  Software design, as a collaborative and creative activity, provides a valuable context for exploring the influence of LLMs on creativity. This study investigates how and where creativity naturally emerges when software designers collaborate with an LLM during a design task. In a laboratory setting simulating a workplace environment, 18 pairs of software professionals with design experience were asked to complete a design task. Each pair had 90 minutes to produce a software design based on a set of requirements, with optional access to a custom LLM interface. Pairs were not primed to be creative. We find that creativity was present in all pairs in design processes, with 13 producing design documents containing creativity. We primarily attribute creativity to the human designers, driven by traits such as prior experience, empathy, and the use of analogies. The LLM contributed by producing novel ideas and elaborating human ideas. However, in some cases, the LLM appeared to hinder creativity by suggesting complex solutions or adding to unproductive digressions. LLMs can support creativity in collaborative software design, but human insights remain central. To effectively augment human creativity, designers must be intentional in their engagement with LLMs.
\end{abstract}


\maketitle

\section{Introduction}
Large Language Models (LLMs) have been adopted broadly by software engineers due to their ability to speed up and improve various aspects of software engineering (SE) activities.
To date, efforts have mainly focused on using LLMs for software development and maintenance tasks, with the majority addressing generation and classification problems \cite{hou_slr24}.
Results indicate clear benefits of LLMs, e.g., in developer productivity \cite{weiszImpact2025,nikolov2025google}.

In many SE-related tasks, such as writing code, testing, or designing, software engineers engage in creative problem solving activities \cite{brooksdesign}.
That is, they do not only seek functional solutions, but additionally \textit{ideas or artifacts that are new, surprising, and valuable} \cite{boden_creative_2004}.
Thus, this focus on creative problem solving can have an important impact on contributing novel and valuable features to software products \cite{maiden2007can,maiden2010requirements}.
As a result, substantial research has focused on creativity in various aspects of SE (e.g., \cite{petre_contrasting,mohanani_requirements_2021,jolak_design_creativity,groeneveld_exploring_2021,gamaStartupsCreativity2025}).

While the potential of using LLMs for creative activities has been picked up in existing research (e.g., \cite{shaer_brainwriting,ivcevic2024artificial,grilli2024creativity}), only a few recent studies focus on this topic within SE (e.g., \cite{wei_ai_ui_design,maiden_genai,inman_seamful_ai,falk2025hackathons}).
This motivates a need for further research on using LLMs for creativity in SE \cite{jackson_genai_creativity}.

Representing an important part of SE practice, collaborative activities are commonplace \cite{herbsleb2007global}.
Existing research has shown how LLMs can support collaborative work in other domains, including for creative activities \cite{liuCognitiveStylesDesign2024,he_group_ideation_llm,yuanning_humanai_collaboration}.
However, integrating LLMs in collaborative, creative SE activities has not yet been studied.
%
%

%
%

To address this gap, we investigate how and where creativity emerges \textit{naturally} in collaborative software design sessions involving pairs of software professionals with access to an LLM. 
Our research question (RQ) is thus: \textit{How, and where, does creativity appear naturally when designing with an LLM?} We conducted a laboratory study simulating a remote workplace environment with 18 pairs of software professionals.
Pairs produced software designs based on an initial set of requirements and had access to an LLM interface that they could choose to use should they wish to.
Creativity appeared in the process of all 18 pairs, which ultimately led to 13 of the pairs producing a design document that contained creative elements, showing that creativity emerges naturally, even when pairs are not primed to be creative, e.g., as in hackathons \cite{falk2025hackathons}.
%
We attribute this creativity primarily to the software engineers, though
the LLM contributed by producing some novel ideas and by elaborating on the input provided by the engineers.
Notably, we observe that LLM use hindered creativity in several cases as well.





\section{Background and Related Work} \label{related_work}

\subsection{Creativity Definitions}
While there are many definitions of creativity~\cite{runco_standard_2012}, we use Boden's definition~\cite{boden_creative_2004}, \say{Creativity is the ability to come up with ideas or artifacts that are new, surprising and valuable}. This definition reflects the creative process \textit{and} the resulting product, recognizing that creative (new, surprising, valuable) products or features do not emerge without some sort of creative activities feeding it. Our research seeks to understand both, and particularly uses Boden's perspective of new, surprising, and valuable as the lens to identify creativity in the product (software design).

To identify creativity in the design process, our study uses Dorst and Cross's concept of a \say{creative event} as an analytical lens. In studying industrial designers, they observed that designers would commence exploring the problem space, frame a small part of the problem space and, from this, make a leap to a small part of the solution space to extend it in some way before returning to the problem space and further considering the problem. Thus, the problem and solution spaces co-evolve~\cite{Maher1996}. When a leap is more than a routine back-and-forth, and leads to the emergence of a novel concept that is visible in the resulting design, 
the moment of insight when the problem and solution spaces are framed leading to the identification of the bridge is considered by Dorst and Cross as a \say{creative event}~\cite{cross_creative_leap_2025}. This concept of a creative event has been used to explore creativity in software design previously~\cite{jolak_design_creativity} and is the one we adopt in this study also (see~\autoref{method}).


Creative expression and experiences vary across individuals. Recognizing that individual creativity exists on a continuum, Kaufman and Beghetto categorize creativity on four levels in their 4C model~\cite{kaufman_beyond_2009}. \say{mini-c} is personal creativity that is meaningful to the individual, \say{little-c} is everyday creativity often used in problem solving and perhaps only recognized by the individual, \say{pro-C} is professional creativity requiring domain expertise and recognized by others, and \say{Big-C} creativity is groundbreaking leading to major innovations in a field. We refer to the 4C model in the final discussion to contextualize some of our findings.


\subsection{Creativity in Software Design}
Although software design has long been considered creative~\cite{brooksdesign}, there is little research on creativity within software design. 
Two papers~\cite{mohanani_requirements_2021, mohananiHowTemplatedRequirements2022} illustrate how framing requirements to give the impression of certainty inhibits the creativity of designers tasked with producing a User Interface design concept. The first~\cite{mohanani_requirements_2021} finds that prioritized requirements led to less original but more practical designs than those resulting from requirements framed as ideas. The second~\cite{mohananiHowTemplatedRequirements2022} 
finds that fixation and reduced critical thinking ensue when designers are presented with more formalized requirements. The combination of the findings of the two papers suggests that, if more innovative solutions are needed, requirements should be framed as ideas, with the resulting ambiguity forcing designers to explore both the problem and solution space. 

In a study examining the effect of physical distance between partners on creativity in the software design process, Jolak et al.~\cite{jolak_design_creativity} find that distance does not make a difference to creativity. 
Specifically, the authors use creative events (as described above) to analyze creativity in the design processes of pairs of software designers, and find no difference in the presence and number of creative events between co-located and remote pairs.  While our study also uses creative events as an analytic lens into creativity in the design process, it differs in that it also examines creativity in the creative product (the software design).

Finally, in reflecting on three decades of research on high-performing software teams, it was noted that experienced designers use contrasts and specific design moves to spark creativity in software development~\cite{petre_contrasting}. An example of such a design move is to identify structural resonances between the problem in hand and similar problems or solutions. Moreover, experienced designers exhibit a design creativity mindset, such as practicing reflection. 

Our study complements these prior studies on software design by making an LLM available to designers, thus enabling an exploration of how creativity emerges with and without the help of the LLM.

\subsection{Generative AI and Creativity}
Prior studies on GenAI in various software development contexts have found that LLMs are not particularly used for creative tasks \textit{in practice}. A study on creativity in start-ups found that LLMs had limited use for creative work~\cite{gamaStartupsCreativity2025}, while another on the adoption of LLMs by software teams notes that GenAI was little used in tasks often considered as requiring creativity, such as design~\cite{pereira2025exploring}. 

In contrast, several studies have shown how LLMs, and GenAI more broadly, could \textit{in theory} assist with creativity in software engineering. 
One study explores the use of GenAI to trigger creativity within application designers~\cite{wei_ai_ui_design} when designing new user interfaces. A second makes use of an approach that combines Natural Language Processing techniques and an LLM to identify novel candidate product features from existing project information~\cite{maiden_genai}, while a third argues that introducing \say{seams} into the user interfaces of AI-powered developer tools can foster creativity by exposing underlying complexity rather than hiding it (and thus forcing developers to consider the problem as much as the solution). We contend that additional approaches need to be explored before results may trickle to practice.

Speculating that creativity will be more important than ever in a world dominated by GenAI fuels a broader research agenda for exploring the impact of GenAI on creativity within software engineering~\cite{jackson_genai_creativity}. The research agenda identifies five themes for future research: creativity as it pertains to: (1) the individual, (2) team, (3) product, (4) unintended consequences, and (5)  wider society. By exploring the influence of an LLM on a collaborative task (software design), our study thus contributes to the team and product themes.

Beyond software engineering, researchers have studied how users can co-create with GenAI on various creative tasks, such as writing~\cite{kim2023metaphorian}, music composition~\cite{louie2020novice}, and meme generation~\cite {wu_llm_memes}. Within this body of research, one stream of inquiry examines whether humans ideating with GenAI generate more novel ideas compared to human ideation alone. For example, one study explored the use of a GenAI-powered design tool that provides inspirational images~\cite{kim2023effect} to aid in a design task, finding that users of the tool produced more novel and varied ideas compared to those who used a version that provided random images. Another study explored the use of an LLM in group brain-writing tasks~\cite{shaer_brainwriting} and noted that the LLM helped support divergent thinking, although there was some evidence that the ideas were uncreative. 

A related focus area is exploring how groups can effectively incorporate GenAI-based tools into collaborative creative work involving groups of humans and GenAI (e.g, ~\cite{he_group_ideation_llm, yuanning_humanai_collaboration, houde2025controlling}). Benefits of such tools include that they can help ideas to be quickly tested and discussed in the group~\cite{yuanning_humanai_collaboration}, maintain the flow of ideas if the group experiences a lull in ideation~\cite{he_group_ideation_llm}, and validate human-generated ideas~\cite{houde2025controlling}. Some challenges of using such tools are that they can stifle human creativity~\cite{houde2025controlling}, the potential for homogeneity in the generated ideas~\cite{he_group_ideation_llm}, and a lack of trust in the answers~\cite{yuanning_humanai_collaboration}.

\section{Research Method} \label{method}
To answer our research question of \textit{``How, and where, does creativity appear naturally when designing with an LLM?''}, we decided upon a laboratory-based setting that purposefully simulates a remote working environment. Moreover, we decided to recruit software professionals with some experience in software design. This decision recognizes that software design is often carried out by people in varying roles rather than being the sole preserve of a full-time architect or software designer. Finally, participants had access to an LLM, but were not required to use it. This approach was taken to allow participants to work in a way natural to them, further simulating a workplace environment. 
The materials provided to participants, coding guidance, and coding examples are available~\cite{supplementary_data}.

\subsection{Study Design}
Our exploratory study 
simulates a remote working environment where professionals collaborate in pairs to complete a design task. The study was designed not to be prescriptive, allowing participants to complete the task as they would in a typical work environment. As we wished to see how creativity naturally arose, we did not ask the participants to \say{be creative}. Instead, they were told the design would be judged on completeness.

The task asked pairs of participants to produce a design document that satisfied a set of product requirements provided in a Product Requirements Document (PRD). 
The PRD contained a list of features and high-fidelity mock-ups for a bicycle parking application aimed at university students seeking secure bicycle parking spots on or near campus. The features were: (1) find bike parks near their current location, (2) search for a destination to find nearby bike parks, (3) get directions to the bike park, (4) pin the bike park to indicate where they park their bicycle, (5) read reviews of a bike park, and (6) crowdsourced bike park reviews. 

A browser-based chatbot in the form of a custom LLM wrapper was developed that enabled interaction with an LLM (specifically OpenAI's ChatGPT 3.5 Turbo). The LLM wrapper logged all prompts and responses entered by participants. The LLM wrapper was accessible to a single user and made available to all participants.

Note, this paper complements another paper~\cite{jackson2026_collab_design} that uses the same dataset to examine how professionals incorporate an LLM into their software design process. 

\subsection{Participants} \label{participants}
Eighteen pairs of software professionals based in the U.S. participated in the study ($n=36$ individuals). Recruitment was conducted through directly contacting potential participants via the researchers' professional networks and a LinkedIn post. To further mimic a workplace setting where teammates collaborate on a task, each potential participant was asked to identify and recruit their own partner. In two cases where participants were unable to find partners, the researchers matched pairs. The demographics of the participants varied (see supplementary data~\cite{supplementary_data}). Most ($n=25$) had used an LLM at work, while nine had used an LLM for personal use. 27 of the participants identified as male and nine as female. The average years of professional work experience was 11 years, with 13 participants having less than 5 years of experience, and 7 participants having 20 years or more of experience. All self-identified as having some experience designing software, with the majority ($n=25$) working in technical roles (e.g., engineer, architect), and others working in roles such as product management. We thus use the term 'designers' in this paper to acknowledge that all participants have some design experience. We refer to specific participants as $P_{xa}$ and $P_{xb}$ where $x$ refers to the pair, $a$ is one partner, and $b$ the other partner (e.g., $P_{6a}$, $P_{6b}$). 

\subsection{Procedure}
All sessions were conducted via Zoom. Participants were given 90 minutes to complete the design task. Except for P8, who worked in the same physical space, all pairs worked remotely from separate locations. Participants were permitted to use any tools and create any artifacts they deemed necessary to articulate their design. Following the task completion, a short exit interview was conducted to understand their experiences.

During the task, participants had access to the custom LLM wrapper. Participants were informed of the tool's availability but were not required to use it. Two pairs (P3, P11) chose not to use it.

Task completion times varied. Eight participants completed the design task before the 90-minute deadline. The shortest completion time was 59 minutes and the average was 82 minutes. The sessions were performed between October 2023 and March 2024.

\subsection{Data Collection}
For each pair, we collected the design session recording (including the interview), their design document, and the LLM wrapper logs. Participant demographic information was collected via an online survey. A professional transcription agency transcribed all recordings of the design task, while the interviews were auto-transcribed by Zoom with subsequent review and correction by a researcher. 

\subsection{Data Analysis} \label{data_analysis}
To answer the research question, we conducted a qualitative analysis of creativity in both the design process and the product (the design document). We decided to identify all instances of creativity in the process and the product, regardless of whether the LLM was involved. This approach facilitates a comparison between creativity arising with and without LLM use. 
Two researchers with expertise in software design undertook the data analysis. One researcher has significant experience building software applications in industry, while the second has prior experience analyzing design conversations for creativity. To aid consistency in coding, we prioritized negotiated agreement between the two researchers over the computation of Inter-Rater Reliability (IRR). Below, we describe our analysis with further details available (see ~\cite{supplementary_data}).  

\subsubsection{Creativity in the Design Process}
We operationalized creativity in the design process as Jolak et al.~\cite{jolak_design_creativity} did, using Dorst and Cross's definition of creative events (described above) to identify creativity in the software design process.
As per Jolak et al.~\cite{jolak_design_creativity}, \textit{a creative event} was identified if a snippet of the designers' conversation could be considered related to the identification of: (1) a connection that simplifies an issue resulting in part of a solution to the problem, (2) an issue that complicates the design problem, (3) an appealing simple solution to an issue, (4) a sudden realization (an \say{aha} moment) related to the problem or solution, or (5) an understanding of the problem.
This definition served as the starting point for the codebook used to abductively code the design session transcripts. 

The two researchers independently coded the first three transcripts before discussion, leading to the identification of additional codes for the code book: \say{analogies}, \say{LLM aiding problem}, \say{LLM aiding solution}, and \say{idea of LLM use}. This revised codebook was re-applied to the first three transcripts. Subsequently, one researcher coded the odd-numbered transcripts, and the second coded the even-numbered transcripts. The researchers then reviewed each other's coding and discussed any disagreements to reach consensus across all transcripts. No new codes were identified in this final round of coding.

Upon completion, the lead researcher reviewed the entire set of creative events across all transcripts to assess their commonality. 
Several themes were identified, including \say{re-use of third-party libraries}, \say{deferring design}, \say{analogies facilitating insights}, and \say{missed opportunities}. 
Additionally, the creative events resulting from LLM use were reviewed closely to determine what influence the LLM had on creative events, leading to several themes, including \say{identify simple solution}, \say{unveil issue(s)}, \say{decision making}, and \say{hindrances}. These themes, along with examples of creative events, are presented in~\autoref{creativity_design_process}.

\subsubsection{Creativity in the Product (Design)} To understand how and where creativity arises in the product (the software design), we needed to: (1) identify designs that were creative and (2) the source of this creativity (the human or the LLM). We decided to use Boden's definition of creativity \say{...artifacts that are new, surprising, and valuable}~\cite{boden_creative_2004}. To operationalize this, we considered a design as creative if it contained \textit{at least one element} that was \textit{new AND surprising AND valuable} in part of any of the different kinds of artifacts that the design contained
(e.g., data models, architectural components, API definitions), since creativity could arise in any of those.
Because the PRD provided a concrete framing of the problem, it served as the baseline against which all designs were considered. Specifically, an element was considered \textit{new} if it did not appear in the PRD and could not be reasonably inferred as an implicit requirement. Routine features (e.g., user login/registration, the reuse of OAuth (P1) or Google Services (P17)) were treated as conventional features, whereas bike-park occupancy tracking (e.g., P8) were considered new as the PRD made no mention of such capabilities. An element was considered \textit{surprising} if it was not considered predictable from the requirements. For example, sourcing external police theft records (P3) and introducing a lewd image detector (P5) were considered surprising as these extend beyond the problem space. Minor variations (filtering results similar to Google Maps (P3)) were not considered surprising. An element was considered \textit{valuable} if it extends the problem space in a way that aligns with the goals of the design problem, or contributes a helpful solution. Examples of valuable additions included non-functional architectural concerns to improve the scalability or performance (e.g., P15, P18) and generalizations to improve adaptability of the system (P10). 
Although Boden’s definition does not explicitly mention usefulness, we excluded features that added undue complexity without a clear benefit. This aligns with the practical meaning of creativity in software engineering, where usefulness is considered integral~\cite{inman_developer_2024}. However, none of the final designs contained complex elements, as designers either did not consider them or ruled them out themselves. For example, P18 ignored the LLMs suggested architecture that combined capabilities from multiple cloud providers, instead proposing an architecture from a single provider. 



\paragraph{Identifying creative elements} The two researchers reviewed one design together (P2) to identify creative elements (e.g., \say{metrics to track usage}) and discussed any issues encountered in using the definition detailed above. Next, both researchers independently identified creative elements in the remaining 17 design documents. They then cross-checked each other's work and discussed any differences to reach full agreement on the creative elements across all design documents. Such an agreement mitigates the subjectivity inherent in judging for newness, surprise, and value.

\begin{table}
    \centering    
    \begin{tabular}{p{0.02\linewidth}p{0.25\linewidth}p{0.6\linewidth}}
    \toprule
       \textbf{\#} & \textbf{Source}  & \textbf{Description}\\
       \midrule
        1 & Knowledge/Exp. & A designer's prior knowledge or experience provides an idea\\        
        2 & Empathy & A designer imagines using the application themselves (literally or ``as someone else'') \\       
        3 & Ambiguity & An ambiguity in the requirements led to varying interpretations\\   
        4 & Analogy & Discussing or exploring analogical applications gave insight into a novel feature or solution\\        
        5 & Underlying goal & Considering the underlying root problem of the design task or the goal rather than the requirements directly\\
        \midrule
        6 & LLM sparks & Unexpected insight triggered in a human who is reviewing LLM output\\
        7 & LLM elaborates & Human-initiated idea is enriched by the LLM\\
        \midrule
        8 & LLM provides & LLM-suggested idea not considered previously by the designers\\
        \bottomrule
    \end{tabular}    
    \caption{Eight sources of creativity in the product. \#1-5 are human-initiated, \#6-7 are the result of a human-LLM collaboration, \#8 is suggested directly by the LLM.}
    \label{tab:creativity_sources}
\end{table}

\paragraph{Source of creative elements} To identify the source (human or LLM), each creative element was triangulated with the corresponding transcript and the LLM wrapper logs. The lead researcher conducted the analysis, while the second reviewed the results; any disagreements were discussed. The triangulation identified the creative events, conversation snippet, and, if the use of the LLM was observed, the LLM prompts and responses leading up to the identified creative element. Pertinent information was captured in a memo. On reviewing the memos, eight common sources for the creative elements (see~\autoref{tab:creativity_sources}) were identified: five human-initiated sources and three involving the LLM in different ways (LLM sparks, LLM elaborates, LLM provides). 
Finally, across all designs, the different sources were counted to assess their prevalence. 

\subsection{Ethics}
The study adhered to the Human Research Protection protocols from the University of California, Irvine (IRB: \#3652). Each participant was provided with a study information sheet ahead of the study and gave verbal consent at the start of the task. 



\section{Findings}
This section presents our findings to the research question: \say{\textit{How, and where, does creativity appear naturally when designing with an LLM?}} We include within our findings all observations related to the appearance of creativity, in both the process and product, not just where creativity arose due to the use of the LLM    . 

\subsection{Summary of LLM Usage}\label{summary_llm_usage}
To provide context for the findings, we note the varied use of the LLM by the 18 pairs. Two pairs (P3, P11) \textit{did not use} the LLM wrapper to assist in their design. P3 preferred to solve the design problem themselves, and P11 did not trust LLMs more generally and consciously decided not to engage with it. Three pairs (P7, P9, P14) \textit{used the LLM to produce their entire design} through ongoing prompting of the LLM wrapper. Five pairs (P4, P5, P10, P15, P17) used it to \textit{seek information} helpful to their design, while creating the design themselves. For example, P5 used the LLM wrapper to explore the Google Maps API, which informed their design. Eight pairs (P1, P2, P6, P8, P12, P13, P16, P18) used it to \textit{seek information and generate parts of the design}, which they subsequently reviewed and incorporated into their design. For example, P1 asked the LLM wrapper to create a data model that they then refined by hand.


\subsection{Creativity in the Design Process} \label{creativity_design_process}

Across \textbf{all 18 pairs, process-level creativity was evident}. Pairs often achieved creative leaps by simplifying (assumptions, re-use, deferrals) and by analogical transfer from similar applications. LLMs rarely reshaped the problem space. Instead, their influence was on solution proposals. Sometimes the LLM was helpful, other times it anchored decisions or provoked unproductive digressions. Before discussing these in greater detail, we first present three vignettes of creative events from three different pairs to illustrate how creativity emerged in the design process. Note, words related to the qualitative coding of the creative events are shown in quotes (e.g., \say{complication}, \say{realization}) in the vignettes.

\subsubsection{Creativity in Action}\label{creative_event_vignettes} 


The first example illustrates the role of re-use in finding an appealing solution to a complication that had arisen. P1a raised a \say{complication} with the design problem in that the requirements do not mention user registration, despite the pair deciding earlier that user accounts were required to store bike park reviews. In parallel, P1b was amending the proposed database design. In so doing, P1b noticed the database design mentioned users, and this caused the \say{realization} of needing user registration functionality, before suggesting a \say{simple solution} that re-used a third-party component \myparticipantquote{OAuth}{P1b} for user login. This \say{realization} and suggestion of a \say{simple solution} helped them make the creative leap from the problem to the solution.


P13 similarly preferred a \say{simpler solution} at one point in the design process. Specifically, at an earlier point, they decided a user should have multiple pins to indicate where their bike was stored or a preferred bike park. When designing the data model for the user, they discussed this further, and simplified their solution by deciding to make an \say{assumption} about the requirements as they designed the data model, which led them to decide on one pin only as \myparticipantquote{That's easier to model}{P13a}. 
This \say{simplification} helped them make the creative leap from the design problem (the need to store location) to the solution (a pin). 


In our final example, inspecting a part of a solution generated by the LLM led a pair to identify a \say{complication} in the LLM's provided design, which they resolved by exploring an \say{analogical} application that provided an \say{insight} into a simplified solution. Specifically, P2  used the LLM wrapper to define a database schema. In reviewing the tables in the LLM's response, they noted it had a separate \textit{LocationSearch} table that stored each search as a location. This led P2a to identify a \say{complication} about what should be stored in the user's search history: the final location selected or the search term itself. To resolve the complication, they opened an analogous application (\textit{Google Maps}) to see how it handled search history, before deciding to adopt its simple solution of storing the search term. The LLM provided design was subsequently amended to store the search term. 
The \say{insight} helped them make the creative leap.

\subsubsection{Insights through Simplification} \label{dc_rq2_simplifying_findings}
Often, a simplification from the pair themselves helped identify the creative leap needed to resolve a dissonance between the problem and the solution. Mechanisms used for simplification included making assumptions or straight-up decisions to go in one direction or another, re-using existing solutions, and deferring part of the design to later.

\textbf{Assumptions and decisions shaped the space} of the problem or solution. This shaping made it easier to identify potential solutions for part of the problem and to take the creative leap necessary to bridge the problem and solution spaces.
Sometimes, the creative leap was made by making an assumption about the problem leading to a reduced scope and simpler design, \say{\textit{Is this app going to be U.S. only? In which case, we’re just defining address in the U.S. format.}}(P7a). They thus designed their database table to support U.S. addresses only.
At other times, decisions about the solution helped with the creative leap. For example, P14 decided the application would be \say{\textit{agnostic to device, so iOS or Android}}, settling on a technology that supported both operating systems (React-Native). 

Other pairs \textbf{simplified their solution by using third party libraries} for addressing part of the design problem. This re-use made it easier to take the creative leap. Such re-use has the dual benefit of both simplifying the solution, as they can rely on the third party to provide part of the solution, and reducing the amount of work for the designers. For example, P17, who, on determining that user accounts would be required for user reviews, decided to use Google authentication services, \say{\textit{So I think using Google or whatever, email to log in and not store the credentials is critical}}. 

Another mechanism employed by the pairs to simplify the design \textbf{was to defer part of the design} until it was truly required to support the application's user base. Such deferring simplified the solution space. As an example, some pairs discussed the initial size of the user base and decided components to support scalability and performance could be deferred, \say{\textit{Let’s focus on the small first. Cause it sounds like this is an initial product. So there’s no need to build something that’s like super scaled.}} (P2). 

Shaping the problem and solution space through assumptions, decisions, clever reuse, and active deferral simplified the design task to help the designers take the creative leap necessary to progress their design.

\subsubsection{Use of Analogies} 
There was widespread (14/18 pairs) use of analogies when discussing the design. This use of \textbf{analogies helped to identify appealing solutions to problems}. 
Mobile applications (e.g., Google Maps, Yelp, Uber) with similar features, such as searching for a place, getting directions, and reviews, were often mentioned. 
Indeed, 6/14 pairs mentioned Yelp in the first five minutes of their design session, sometimes influencing their future design. At other times, analogies were used to discuss a specific aspect of the design problem. For example, in discussing how users may want to use review criteria to help identify a bike park, P3b described how Google Maps lets one filter on certain criteria and suggested their app could do the same. 

Some pairs went beyond just mentioning an analogy, and turned to engaging with analogical applications in detail to see how they solved a problem similar to the one they faced. 
For example, when modeling their database, P1 were unsure whether photos should be linked to the review of a bike park, directly to the bike park, or both. Therefore, they opened up Google Maps to explore how it supported photos for a location. Doing so, they decided to limit the scope, thus shaping the problem space, \say{\textit{There’s no requirement for users to submit their own bike park locations.}} (P1b) and simplifying their solution by deciding, \say{\textit{users can only submit photos as part of a review}} (P1b). 


Analogies helped address issues or complications, thus facilitating creative leaps. This behavior is to be expected, given that designers are known for using analogies when solving problems~\cite{cross_creative_leap_2025}. 

\subsubsection{LLM Influence on Creative Events} \label{dc_llm_involve_ce}


In using the LLM to automate the more mundane aspects of design, such as outputting boilerplate API definitions, \textbf{the LLM provided space for the pairs to think more deeply} about the design problem. Thus, it indirectly influenced creativity, \myparticipantquote{It gives the foundational kind of like busy work out of the way. And so it provides more time for me and [P18a] to have discussions about like, does this make sense like, what, how do we refine this?}{P18b}. Additionally, we observe that \textbf{the LLM primarily assisted with finding a solution rather than reshaping the problem space}. Indeed,
we did not see any instance of the LLM for reshaping the problem space or the LLM itself suggesting considerations, including complications---these parts of a creative event always came from the designers themselves.

As noted in~\autoref{summary_llm_usage}, many pairs asked the LLM to generate part of their design. While not all generated parts were simple, sometimes \textbf{the LLM offered a simple solution} that the pair could immediately use. P14 asked the LLM to define the database tables needed. On reviewing the solution proposed by the LLM containing \say{Locations}, \say{Reviews}, and \say{Users}, they decided those were \say{\textit{the three main things}} and kept them in their design. Other times, the pair asked the LLM for multiple options and, after reviewing, decided which to select. For example, when discussing the backend architecture, P6 asked the LLM for \say{\textit{What are the difference choices for the backend architecture for this app}}, and decided upon serverless. These two cases illustrate contrasting human behavior. In the first, they did not consider alternatives, unlike the second. 
This points to the risk of anchoring to the first proposed solution, which could hinder creativity by preventing exploration of alternatives.

In reviewing LLM responses to prompts, the discussions among the pairs sometimes \textbf{unveiled issues and complications with the problem or solution}, requiring insight from the designers to resolve and make the creative leap. Consider P2, who asked the LLM to generate a set of server APIs. On reviewing the response detailing an API for finding bike parks near the current location, they noted the API represented the location as latitude and longitude. P2 noted this could be problematic as users would be searching by a location, not by latitude and longitude. To resolve the issue, they went back to the PRD and spent some time walking through the user flows before concluding that the provided APIs would still work.

The \textbf{LLM aided in the decision-making} necessary to facilitate the creative leap from the problem to the solution space on several occasions. Sometimes the LLM implicitly made a decision when proposing a design, other times it was the designers who made the decision based on information from the LLM. For example, P7, who used the LLM to generate their entire data model, asked it to include the user's prior searches. It did so as an array of strings, which P7 agreed with on review. This simple solution thus enabled the pair to take the creative leap from the problem (need to store searches) to a solution (updated database design). Note that such straightforward decision-making did not always occur. Indeed, in other cases, the designers had to significantly amend the proposed design after a review, because they were not happy with the design choice made by the LLM (e.g., P2 amended suggested APIs). Moreover, the LLM sometimes did not make a decision at all. When P6 asked the LLM for advice on the best backend architecture, for instance, it explicitly declined to make a choice. Instead, it provided considerations to aid the decision-making and left the decision to the pair. 

Sometimes, the LLM actively \textbf{hindered the designers through over-complications and provoking digressions}. For example, in one case, it produced an overly complex design. P18 asked it to help with the architecture design, but found the response unhelpful as it referenced components from two different cloud providers: AWS Cognito and Google Firebase. Another time, the response caused an unnecessary digression. P7 had already reached consensus on the use of user accounts, but upon receiving a response from the LLM referencing user accounts, the conversation started afresh, leading to an unncessary digression. 

\subsubsection{Missed Opportunities for Creativity in the Product}
In discussing the problem and potential solutions, some designers mentioned novel ideas that would have led to creativity in the product. Yet, while some were elaborated and included in the final design (discussed below), others were not. Sometimes this was due to a conscious decision to declare an idea out of scope. For example, P11 raised the possibility of a \say{\textit{Report closed}} button, inspired by a similar feature in Yelp. However, they dismissed it as \say{feature creep}, and simply noted it as an idea rather than elaborating on it further in their design. Other times, novel ideas were mentioned, yet not discussed further or revisited, even when the pair liked the idea. For example, P18 briefly discussed an appropriateness filter on review comments early in their design, which they agreed would \say{\textit{definitely}} be needed. Yet, it was not mentioned again. 
Responses from the LLM that included novel suggestions were sometimes treated in the same way as well. P10a reacted favorably to the suggestion within their first LLM response that users could share their reviews and favorite bike parks, \myparticipantquote{Social sharing. Oh, that’s good. I like that.}{P10a}, yet this feature was not included in their design.





\subsection{Creativity in the Product} \label{dc_creativity_in_product}
In contrast to the observation that creativity in the process was present for all, fewer pairs had creative elements in their design documents. We elaborate further on product creativity. 

\subsubsection{Presence of Creativity within the Design} \label{creativity_in_design}
\begin{figure}
\centerline{\includegraphics[width=0.9\columnwidth]{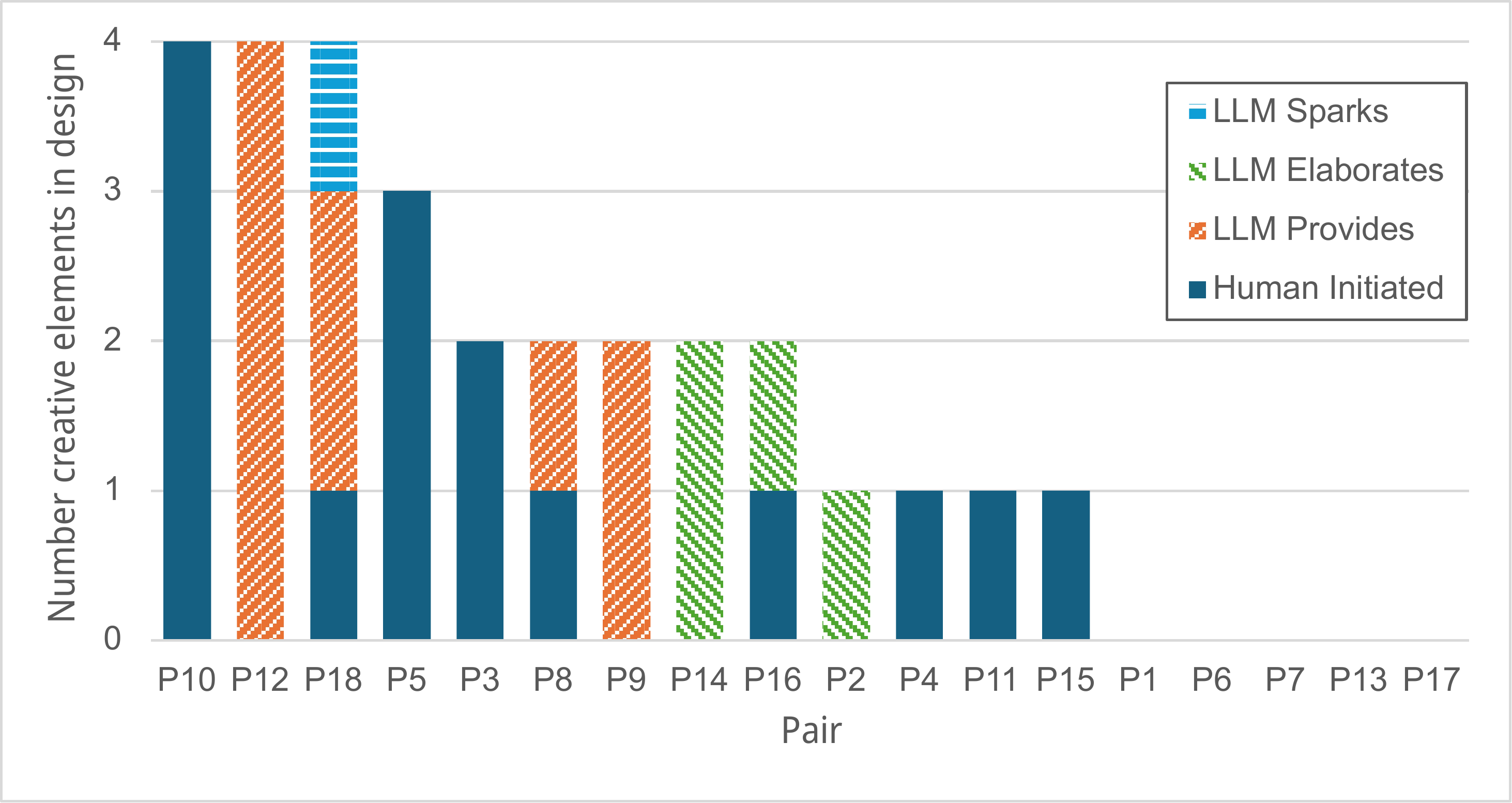}}
\caption{Number of creative elements included in the design document of each pair, categorized into the source (human or LLM). 
}

\label{fig:creative_features_per_pair}
 \Description[Creative Elements in each pair's design documents]{Shows the number of creative elements in each pair's design document. P10, P12, P18, all had four creative elements. P5 had three, P3, P8, P9, P14, P16 all had two. P2, P4, P11, P15 all had one. P1, P6, P7, P13, P17 had no creative elements.} 
\end{figure}

Although the designers were not primed for creativity, \textbf{many design documents (13/18) included at least one new, surprising, and valuable element} (see ~\autoref{fig:creative_features_per_pair}). The number of creative elements varied: three design documents contained four, while five had none. The most common creative element was scalability of performance concerns (five designs), followed by tracking bike park availability (four designs). Differences in the number of creative elements reflect different attitudes to scope expansion and the varying context provided to the LLM.  

\begin{table}
    \centering
        \caption[Creative Elements Presented in the Designs]{Creative elements presented in the designs by pair. If LLM assisted, superscripts are used to indicate how: e for elaborate, s for sparks, p for provides}
\begin{tabular}
{p{0.01\linewidth}p{0.63\linewidth}p{0.25\linewidth}}
\toprule
\textbf{\#}& \textbf{Creative Element} & \textbf{Pairs} \\ \midrule
      1 & Scalability or performance solutions & P8, P14\textsuperscript{e}, P15, P16, P18\textsuperscript{s} \\
      2 & Track bike park capacity and occupancy & P3, P8\textsuperscript{p}, P9\textsuperscript{p}, P12\textsuperscript{p} \\      
      3 & Scanner to detect thefts from police reports & P3, P5 \\
      4 & Data protection requirements & P16\textsuperscript{e}, P18\textsuperscript{p} \\
      5 & Lewd image scanner on image upload & P5 \\
      6 & Profanity checker on review submission & P5 \\
      7 & Validate user input for spam and inappropriate content & P9\textsuperscript{p} \\
      8 & Content moderation & P12\textsuperscript{p} \\     
      9 & Upvote/downvote reviews & P12\textsuperscript{p} \\
      10 & Bike park accessibility information & P12\textsuperscript{p} \\            
      11 & Push notifications to remind user to review bike park & P10 \\  
      12 & Shows location bike parked on home screen & P11 \\
      13 & Tracks recently used bike parks & P18 \\                
      14 & User configurable radius search distance & P10 \\      
      15 & Metrics to track bike park usage & P2\textsuperscript{e} \\    
      16 & Generalization for multiple universities & P10 \\
      17 & Generalization for many review attributes & P10 \\
      18 & Sources bike parks and reviews from external sources & P4 \\    
      19 & Accessible application & P18\textsuperscript{p}\\  
      20 & Cost optimization & P14\textsuperscript{e} \\
         \bottomrule
    \end{tabular}
    \label{tab:dc_product_novel_features}
\end{table}

Twenty different creative elements were identified across the 13 design documents containing creative elements (see~\autoref{tab:dc_product_novel_features}). 
Four creative elements appeared in multiple design documents, indicating that they are not novel across the entire set of designs. For example, four designs incorporated a feature to track a bike park's capacity and measure its occupancy. Other times, the features were creative elements unique to a single design, e.g., push notifications to remind a user to review a bike park.

The most frequent creative element (five designs) provided a scalability or performance solution. The PRD made no mention of user numbers or projected growth. Yet, some pairs were concerned with ensuring the application could support a reasonable number of users, so they designed the architecture to scale and perform well. For example, P15 included a Redis cache for photos to serve them more quickly to users. 

Scalability or performance solutions being the most common creative element is indicative of a factor that potentially explains why some pairs had more creative elements than others: \textbf{some pairs limited the scope of the problem or solution} by more or less sticking to the PRD, while others extrapolated potentially important additional features. Within~\autoref{dc_rq2_simplifying_findings}, performance and scalability considerations were provided as an example of where some pairs decided that the initial number of users would be small, which is clearly opposite to the example just provided of P15. 
This differing behavior highlights the impact that scoping decisions and intentional limits have on the creativity of the end product.

Five out of sixteen designs from the pairs that used the LLM did not contain creative elements. In contrast, designs from the two pairs that did not use the LLM had creative elements. Although this is a small sample, these figures indicate that \textbf{using the LLM does not guarantee creativity} in the product. Other factors for how creativity arises are at play. One potential explanation could be the amount of problem context provided to the LLM. Four out of five pairs that had no creative elements used the LLM to generate some or all of their designs. In doing so, they all provided much context about the problem by copying requirements from the PRD into the LLM prompts. In effect, they defined a fixed problem space to which the LLM obliged by providing a constrained potential solution that matched the PRD very closely, leading to outcomes similar to those noted in the study on fixed requirements hindering creativity~\cite{mohanani_requirements_2021}. 

\subsubsection{Source of the Creativity} \label{dc_product_creativity_source}


\begin{table}[b]
    \centering
    \caption{Sources of the creative elements within the design documents. Each creative element is counted against one source. The sources are shown with the total number of creative elements and the number of attributed participant creative elements in parentheses.}
    \begin{tabular}
{p{0.3\linewidth}p{0.08\linewidth}p{0.48\linewidth}}
    \toprule
        \textbf{Source} & \textbf{Total} & \textbf{Pairs}\\
        \midrule
         Knowledge/Exp. & 8 & P5 (2), P8 (1), P10 (3), P15 (1), P16 (1)\\     Empathy & 2 & P5 (1), P11 (1)\\
         Ambiguity & 2 & P4 (1), P18 (1)\\
         Underlying goal & 2 & P3 (2)\\
         Analogy & 1 & P10 (1)\\      
         \midrule
         LLM elaborates & 4 & P2 (1), P14 (2), P16 (1)\\
         LLM sparks & 1 & P18 (1)\\              
         \midrule
         LLM provides & 9 & P8 (1), P9 (2), P12 (4), P18 (2)\\       
         \bottomrule
    \end{tabular}    
    \label{tab:dc_creative_sources}
\end{table}

Notably, \textbf{the creative influence of the LLM was less than that of human-initiated sources}, such as prior knowledge or empathy from the designers. Of the 29 creative elements in the designs, 15 are attributable to the five sources characterizing designers' behaviors and skills (see \autoref{tab:dc_creative_sources}), a further five are the outcome of the two sources typifying joint human-LLM creativity (LLM elaborates, LLM sparks), and nine are directly suggested by the LLM (LLM provides).


\paragraph{Prior knowledge/experience} The knowledge and experience of the designers is the most frequent human-initiated source, with eight occurrences across five pairs. Examples include P5, who included a \textit{profanity checker} and \textit{lewd image detector} in their solution. In the exit interview, P5a mentioned these ideas came directly from their work experience, \say{\textit{where I work, we have something somewhat similar, just because we handle a lot of images every day.}} Experience in design helped also, as designers often use design principles gained over the years to guide their work~\cite{petreSoftwareDesignDecoded2016}. For example, P10a (an experienced professional) mentioned design principles such as the desire to \say{\textit{generalize}} and the need to consider \say{\textit{scalability}}. These design principles guided the pair to a generalized database design that supported multiple universities, arbitrary review attributes, and user-configurable features such as search radius. 

\paragraph{Empathy} Two pairs imagined using the application and, in doing so, gained insight into beneficial features. For example, P11 discussed what the application should display upon launch. P11a noted that they, \say{\textit{would want something where I'm just like to go to pin}} to easily find their retrieved bike. This led to the idea of adding a link to the pinned bike park on the application's home screen. 



\paragraph{Ambiguities} Two pairs found ambiguities in the PRD 
leading to the identification of creative elements. One ambiguity was the word \say{crowdsourcing} in the context of user reviews. While the intent of this was to indicate users of the application inputting reviews of a bike park, P4b was unsure what it meant, \say{\textit{Is it the ones that the user has saved on this app? Or is it going to source it from everywhere on the internet?}} After a brief discussion, they decided to include both, and so their design included a \say{Review Sourcer} service to find external reviews from Google. 

\paragraph{Underlying goal of the design problem} One pair disregarded the PRD and decided to focus on three core considerations implied in the goal of the problem, \say{\textit{Do you want to break this up into safe, vacant, and close?}} (P3a). This led to much discussion about the best way to determine how safe a bicycle park was, whether it was vacant or occupied, and how to decide which bike parks to show in the search results. Their conversation thus included aspects such as sourcing bicycle theft records from the police, tracking whether a bicycle has been securely retrieved from a bicycle park, and whether bike parks with high theft rates should be excluded from the search results. 
Their resulting database design contained a bike park table with the novel attributes \textit{total\_bike\_count}, \textit{vacant\_count}, \textit{success\_retrieval\_pct}, and \textit{thefts\_per\_month}. 

\paragraph{Analogies} Although, as noted above, discussing analogical applications with similar features helped to solve the design problem, only one pair (P10) used analogies to unearth creative elements. P10 in particular discussed the challenge of getting a user to return to the application to leave a review or pin the bicycle's location after the application launched a third-party mapping application to navigate to a selected bike park. They resolved the challenge by reusing the approach (push notifications to mobile devices) used by Lyft to remind riders to leave a review of their driver. 

\paragraph{LLM elaborates} The LLM helped four pairs elaborate on ideas generated by the designers themselves. Consider P2b, who raised the need to estimate the operational cost of running the application on a cloud-based infrastructure such as Amazon AWS, noting that at work, \say{\textit{We’re pressed all the time to think about cost.}} This led to a discussion on how to monetize the application to pay for its operational costs, whereupon P2a had the idea that the city should pay for the application. P2a felt the app \say{\textit{probably want[s] to collect metrics}} to provide to the city. Subsequently, they asked the LLM wrapper for metrics that they included in their design.

\paragraph{LLM sparks} Human-AI collaboration was also evident when the LLM's response sparked an idea in P18. They asked the LLM to produce a design document with instructions to consider cloud-native technologies and a 3-tier architecture. The response mentioned AWS S3 for photo storage, which sparked the idea to add a Content Distribution Network (CDN) to their design as \myparticipantquote{CDN provides performant access}{P18b}. A CDN was not mentioned in the prior LLM responses or in the conversation, so presumably this decision was based on the designers' prior experience.

\paragraph{LLM produces} While elaborating and sparking can be considered an indirect influence of the LLM on creativity, the LLM also aided more directly by generating a response containing a novel idea not previously considered by the pair. Three pairs were assisted in this way, although their prompting approaches differed. P12's first prompt to the LLM wrapper was broad and open-ended, with no mention of the PRD requirements, \say{\textit{Help me design a mobile app that finds me a secure bike park to park my bike}}. In subsequent exploration of the suggested features through the LLM wrapper, the pair identified novel elements, including content moderation and tracked bike park capacity and occupancy. 
In contrast, P8 and P18 entered prompts requesting specific design artifacts. P8 asked for a data model for a parking lot (an analogy), and the response mentioned capacity and the number of available spots, which they 
included in their design. P18's prompt was more open, giving the LLM flexibility in its response by asking for cross-functional requirements and fitness functions. The LLM suggested accessibility and data protection features, which they incorporated. 
These varying prompting styles are another potential factor in why the LLM's influence on product creativity varies across pairs.

\section{Discussion}
Software design is an integral part of building software applications and often considered a creative task~\cite{brooksdesign}. However, the influence of an LLM on such a collaborative and creative task has been little examined in comparison to studies that consider the use of LLMs in coding~\cite{hou_slr24}. By examining how, and where, creativity naturally emerges when using an LLM in software design, we provide novel insights into the impact of an LLM on creativity in software design. 

We firstly note that \textbf{human creativity remains central to software design}, even when using an LLM. The designers consistently demonstrated creativity in the design process by reshaping the problem, identifying simple solutions that included reuse, and drawing upon analogies to aid problem-solving, regardless of whether they used the LLM. Creativity was present in the design process for all pairs, reinforcing the view that creativity is key to software design~\cite{brooksdesign}.
Notably, this was the case without priming the study participants to be creative.

Creativity in the design process resulted in creative elements in 13/18 pairs' (72\%) designs, with the LLM helping spark, elaborate, and produce creative elements. However, we observe that human traits, such as prior knowledge and empathy, contributed to identifying creative elements more often than the LLM, per~\autoref{tab:dc_creative_sources}, particularly also because both LLM spark and LLM elaborate strongly depend on interplay with humans. 

Our findings also suggest that the \textbf{LLMs' primary influence is on product-level creativity} rather than on process creativity. We observed many creative events~\cite{cross_creative_coevolution_2025} described in~\autoref{creativity_design_process}, with the LLM only playing a role in solution finding rather than reshaping the problem space. Importantly, \textbf{using the LLM did not guarantee creativity}, with 5/16 pairs using the LLM having no creative elements while both non-LLM pairs did, underscoring that availability alone is insufficient without adequate integration into the design process.

While using the LLM led to several creative elements in the sense of novelty, we also observed that \textbf{creative elements attributed to the LLM were not entirely surprising} and did not differ that much from those produced by humans alone (Table~\ref{tab:dc_product_novel_features}). That is, the LLM-produced ideas were new with respect to the PRD, but typically related to commodity functionality, such as adding content moderation. This could be due to our detailed PRD constraining exploration (\cite{mohanani_perceptions_2017}) by the designers, including in their interactions with the LLM. Only one pair using the LLM was observed to have expansive, open-ended prompts devoid of PRD detail, leading to creative elements. This observation encourages a deeper investigation of how LLMs can be used for creative work beyond commodity, perhaps by exploring idea-framed requirements. More broadly, we note the creative elements, irrespective of source (human or LLM), are not at a \say{Big-C} level, as none in~\autoref {tab:dc_creative_sources} are groundbreaking. 


Interestingly, accompanying the more direct influence of the LLM was an indirect effect. Using the \textbf{LLM helped free cognitive space by automating the routine} parts of design, such as writing boilerplate. Such space allowed the pairs to think more deeply about the design problem. This observation suggests that automating rote work through the use of an LLM not only benefits productivity~\cite{ziegler2024measuring}, but can also play a role in aiding creativity. While freeing up of cognitive space through automation has been speculated upon~\cite {jackson_genai_creativity}, our study finds that it can indeed happen. 

Using an LLM was not always positive as, sometimes, the \textbf{LLM introduced distractions}, which could potentially hinder creativity. Additionally, these distractions stand in direct opposition to the previous point, i.e., freeing up cognitive space. One pair in particular spent time on a tangential discussion about a non-essential and well-solved problem (user accounts) as a result of a response from the LLM. This observation highlights the need for a more critical engagement with the LLM to prevent digressions. Continuing to utilize well-understood design practices~\cite{petreSoftwareDesignDecoded2016} can likely enhance such engagement, as can the development of design tools shaped to augment human-led creativity (as discussed below).

We note pairs exhibited differing creativity in their designs, perhaps due to \textbf{varying responses to scope creep}~\cite{elliott2007anything}, irrespective of LLM use. Some pairs seemingly embraced newly-suggested features, while others rejected ideas due to concerns about scope creep. 
The tension between innovation arising from new ideas and the need to deliver has always been present in software teams. With the advent of LLMs and their ability to augment human creativity, teams may need to re-evaluate how they consider scope creep to ensure innovative ideas are not unintentionally squashed.

Another reason some designs lacked creativity could be the varying levels of creativity among our heterogeneous sample. Some have decades of experience and potentially Pro-C creativity, while others are novices likely to be exhibiting mini-c or little-c creativity~\cite{kaufman_beyond_2009}. Moreover, prior research has noted the influence of personality traits on creativity~\cite{amin_personalitytraits_creativity_2020}. Although creativity arose for all pairs in the design process, it did not always lead to novelty in the design product as perceived by external observers. This connects to the concept of everyday creativity~\cite{kaufman_beyond_2009}, which \textbf{exhibits itself as mini-c, little-c, and Pro-c in the creative work} of designers and aids in problem-solving and gaining personal insights. 
Such creativity is important both for completing tasks and for personal reasons such as improved well-being~\cite{smith_creative_2022}, and should not be overlooked as AI is increasingly adopted for software development tasks. 

Finally, our empirical study on where, and how, creativity arises when designing with an LLM, contributes to broader conversations on human-AI co-creation (e.g.,~\cite{ivcevic2024artificial, bouschery2023augmenting, grilli2024creativity}). We have shown that creativity naturally emerges in a collaborative act such as software design, irrespective of whether AI is used. While AI can help in amplifying human creativity, it is not replacing human creativity. This perspective of amplification is shared by HCI scholars such as Shneiderman~\cite{shneidermanHumanCenteredAI2022} and we touch upon it in future work below.

\subsection{Implications for Practitioners}
Our findings lead to several practical implications for practitioners. 
Firstly, when using an LLM, it is important to consider multiple alternatives before selecting a solution. This reduces the risk of anchoring on the first solution that we observed in this study. Secondly, designers should consider open-ended, expansive prompts with minimal context if they wish to elicit creative input from the LLM, rather than providing detailed requirements that constrain it. Interacting in a more playful manner~\cite{Nikghalbplayful2025} can also help to amplify human creativity. Thirdly, designers should perhaps time-box interactions with the LLM, including discussions of its responses, to prevent LLM-induced digressions. Finally, creativity thrives when designers have space to think deeply about a problem, as they can attain the flow state~\cite{ritonummi_flow} known to aid creativity \cite{csikszentmihalyi_creativity_2013}. As noted in the findings, using an LLM to automate the mundane, time-consuming work of design (e.g., completing boilerplate artifacts) can create this space.

\subsection{Future Research Directions} \label{dc_impl_research}

One direction is to explore the influence of an LLM on creativity in an industry setting, as real-world software design is more complex than the task used in this study. 
Such research would help unpack a recent observation that some developers believed ChatGPT aided creativity, while for others it hindered~\cite{meem_WhyUseChatGPT_2025}. Research could also explore whether the \say{space} created by using LLMs to automate rote tasks, really does aid creativity by providing time for deep thinking, or whether it gets swept away as other tasks fill the gap.

A second direction is to explore how novel LLM-based tools can amplify human creativity in software design. Such creativity support tools~\cite{frich2019mapping} could be designed to support practices known to aid creativity, for example by encouraging exploration of the problem space, identifying alternative solution designs, fostering periodic reflection, or mitigating some of the downsides of LLM use in creative work (e.g., design fixation \cite{wadinambiarachchiEffectsGenerativeAI2024}). Ensuring humans retain their agency in creative work should be a key design principle in designing new LLM-based creativity support tools.


\section{Limitations and Threats to Validity}

One limitation is that participants were not explicitly informed that the study would assess creativity. This was by design, as we wanted to see how and where creativity naturally arose in a simulated work environment. 
\say{Big-C} creativity (not observed in this study) may have arisen if the participants had been primed for creativity within the instructions or the requirements were framed as ideas, as noted by Mohanani et al~\cite{mohanani_requirements_2021}. 
A second is that our participants differed in their experience (e.g., years working, role, experience of LLM use). This variety will influence their design processes, their use of the LLM, and the resulting level of creativity. While this heterogeneity is indicative of industry teams, this variety could influence the level of creativity we observed.
Finally, participants had only 90 minutes to complete the task, which could have led to undue time pressure, the impact of which is nuanced with mixed outcomes~\cite{khedhaouriaTimePressureTeam2017, KUUTILA2020106257}. 


The research team has significant experience in software design from research and industry. 
This experience 
facilitated rich discussions when coding and in meetings with the wider team. Moreover, our findings are strengthened by the triangulation of multiple datasets. As this is an exploratory study, we make no claims to data saturation~\cite{saunders_saturation_2018}. Further studies may find alternate ways in which creativity arises in collaborative software design. Findings may not be fully transferable to industry or when designing for an existing application, as the study was laboratory-based. 

To increase the transparency of the findings, we provide a detailed study design  in~\autoref{method} and share supplementary materials~\cite{supplementary_data}. 
The same two researchers performed the data analysis, thus reducing researcher bias. By coding individually, discussing, and reviewing each other's work, we reached consensus on the coding and subsequent findings.  



\section{Conclusion}
Software design is a collaborative and inherently creative activity that forms an essential part of software development. 
Through a laboratory-based study, where 18 pairs of software professionals were asked to complete a design task within 90-minutes and had access to an LLM, we found that human creativity remains a vital part of software design \textit{process}. While the LLM occasionally sparked ideas or suggested novel features, it was the humans who undertook the core creative work of re-shaping the problem, considering scope, drawing analogies with existing systems, and the overall direction of exploration.
We find the LLM's influence was more pronounced---even though still relatively limited---in the design \textit{product}, as a number of designs included novel, surprising, and valuable additions beyond the PRD. That said, the creative elements suggested by the LLM were relatively homogeneous
We recommend that practitioners hone their AI literacy skills in crafting exploratory prompts to better leverage LLMs for idea generation. Moreover, using an LLM to automate routine aspects of design may create space for more in-depth design thinking. Future research directions include exploring how to design LLM-backed creativity support tools to augment human creativity in tasks such as software design.

\begin{acks}
Thanks to our participants for undertaking the design challenge. The authors appreciate Sadid Khan's assistance in developing the chat tool. Prof. Marian Petre from the Open University, UK, provided valuable feedback on the study. Prikladnicki is partially funded by Fapergs and CNPq, Brazil; van der Hoek acknowledges support by the National Science Foundation under grants CCF-2210812 and 2326489.
\end{acks}

\bibliographystyle{ACM-Reference-Format}
\bibliography{refs}

\end{document}